\documentclass[10pt,twocolumn,a4] {article}
\usepackage{graphicx,colordvi}
\usepackage{amssymb}
\usepackage{xcolor}
\title{\bf Fission fragment mass yields of Th to Rf even-even nuclei
\footnote{Supported by the Polish National Science Center (Grant No.
2018/30/Q/ST2/00185) and by the National Natural Science Foundation of China
(Grant No. 11961131010 and 11790325).}}

\author{Krzysztof Pomorski$^1$\footnote{Email Krzysztof.Pomorski@umcs.pl}, 
 Jos\'e M. Blanco$^1$, Pavel V. Kostryukov$^1$, Artur Dobrowolski$^1$,\\ 
Bo\.zena Nerlo-Pomorska$^1$, Micha{\l} Warda$^1$, 
Zhigang Xiao$^2$\footnote{Email: xiaozg@mail.tsinghua.edu.cn}, 
Yongjing Chen$^3$, Lile Liu$^3$,\\ Jun-Long Tian$^4$, Xinyue Diao$^2$, 
Qianghua Wu$^2$\\
{\footnotesize $^1$ Institute of Physics, Maria Curie Sk{\l}odowska University,
     20-031 Lublin, Poland}\\
{\footnotesize $^2$ Department of Physics, Tsinghua University, Beijing 100084,
 China}\\
{\footnotesize $^3$ China Institute of Atomic Energy, Beijing 102413, China}\\
{\footnotesize $^4$ School of Physics and Electrical Engineering, Anyang Normal
 University, Anyang 455000, China}
}
\date{\today}
\begin{document}
\onecolumn

\maketitle

\begin{center} \begin{minipage}{16cm} {\bf Abstract:}  

Fission properties of the actinide nuclei are deduced from theoretical analysis.
We investigate potential energy surfaces and fission barriers and predict the
fission fragment mass-yields of actinide isotopes. The results are compared with
experimental data where available. The calculations were performed in the
macroscopic-microscopic approximation with the Lublin-Strasbourg Drop (LSD) for
the macroscopic part and the microscopic energy corrections were evaluated in
the Yukawa-folded potential. The Fourier nuclear shape parametrization is used
to describe the nuclear shape, including the non-axial degree of freedom. The
fission fragment mass-yields of considered nuclei are evaluated within a 3D
collective model using the Born-Oppenheimer approximation.

 
\end{minipage}
\end{center}
\twocolumn
\normalsize

\section{Introduction}

Good reproduction of fission barrier heights and fission fragments mass-yields is a test of the theoretical models describing the nuclear fission process. An interesting review of the existing fission models can be found in Refs. \cite{MJV19,RMo13,SJA16}. Extended calculations of the fission barrier heights can be found in Refs.~\cite{BKR15,JKS20}. Readers who are interested in the 
theory of nuclear fission can find more details in the textbook \cite{KPo12}.

In the present paper, the fission fragment mass yields (FMY) are obtained by an
approximate solution of the eigenproblem of a three-dimensional collective
Hamiltonian, of which the coordinates correspond to the fission, neck, and
mass-asymmetry modes. Here presented model is described in details in
Refs.~\cite{PIN17,PNB18,PDH20}. The potential energy surfaces (PES) of
fissioning nuclei obtained by the macroscopic-microscopic (mac-mic) method in
which the Lublin-Strasbourg Drop (LSD) model \cite{PDu03} has been used for the
macroscopic part of the energy, while the microscopic shell and pairing
corrections are evaluated using single-particle levels of the Yukawa-folded (YF)
mean-field potential \cite{DNi76,DPB16}. The Fourier parametrization is used to
describe shapes of fissioning nuclei \cite{PNB15,SPN17}. It is shown in
Ref.~\cite{Bar20} that this parametrization describes very well the shapes of
the nuclei even close to the scission configuration. 

The paper is organized in the following way. In Section 2, we present first
the details of the shape parametrization and the theoretical model.
Then we show the collective potential energy surface evaluated within the
mac-mic model for the selected isotopes and our estimates of the fission barrier
heights. The calculated FMY are compared with the existing experimental data in
Section 3. The estimates of FMY for Th isotopes and their dependence on two
adjustable parameters are further discussed in details in Section 4. Conclusions
and perspectives of further investigations are presented in Section 5.

\section{Model of the fission dynamics}

The evolution of a nucleus from the equilibrium state towards fission is
described here by a simple dynamical approach based on the PES. We assume that
at large deformations, the shape of the nucleus should depend on three
collective degrees of freedom describing its elongation, left-right asymmetry,
and the neck-size. At smaller deformations, up to the second saddle, the
non-axial  shapes are also considered. In the following subsection, we present
shortly a Fourier type parametrization of the nuclear shape which is used in the
paper.
\begin{figure}
\includegraphics[width=0.95\columnwidth]{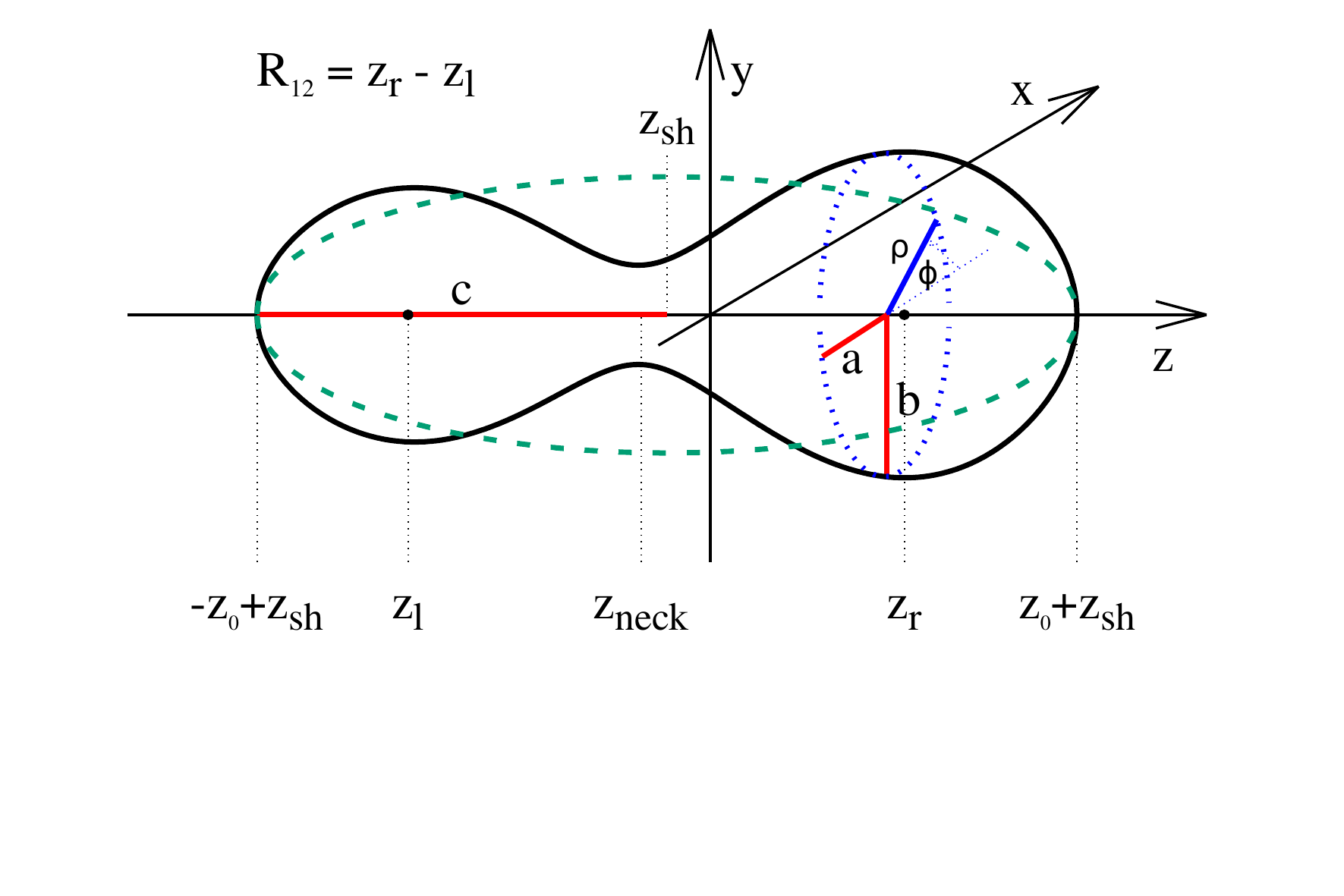}\\[-8ex]
\caption{Shape of a very elongated fissioning nucleus.}
\label{shape}
\end{figure}


\subsection{Fourier nuclear shape parametrization}

A typical shape of the nucleus on the way from the saddle to the scission
configuration is shown in Fig.~\ref{shape}, where $\rho(z)$, the distance from
the z-axis to the surface of the nucleus as a function of $z$, is plotted. Here
by the nuclear surface is treated the surface of the nuclear liquid drop, or the
half-density surface when the microscopic density distribution is considered.

The function $\rho (z)$ corresponding to the nuclear surface can be expanded
in the Fourier series in the following way \cite{PNB15}: 
\begin{equation}
\begin{array}{ll}
 \rho_s^2(z)\!=\!R_0^2\! \sum\limits_{n=1}^\infty &\left[
 a_{2n} \cos\left(\frac{(2n-1) \pi}{2} \, \frac{z-z_{sh}}{z_0}\right)\right.\\
 &+\left. a_{2n+1} \sin\left(\frac{2 n \pi}{2} \, \frac{z-z_{sh}}{z_0}\right)
 \right]~.
\end{array}
\end{equation} 
Here $z_0$ is the half-length of the total elongation of the nucleus and 
$z_{\rm sh}$ locates the center of mass of the nucleus at the origin of the 
coordinate system. The expansion parameters $a_i$ can serve as parameters
describing the shape of nucleus. The length parameter $c=z_0/R_0$ is fixed by 
the volume conservation condition, where $R_0$ is the radius of spherical nucleus having the same volume as the deformed one.

Contrary to frequently used spherical harmonics expansion (conf. e.g., Refs.
\cite{BKR15,JKS20}), the Fourier series converges much earlier for the realistic shape of nuclei \cite{PNB15,SPN17} and only a few first terms are sufficient in
practical use. Although one can work directly with these Fourier expansion 
coefficients treating them as free deformation parameters, it is more suitable to use their combinations $\{q_n\}$, called {\it optimal
coordinates} \cite{SPN17}, as following
\begin{equation}
\left\{
\begin{array}{l}
 q_2=a_2^{(0)}/a_2 - a_2/a_2^{(0)} \\[1ex]
 q_3=a_3 \\[1ex]
 q_4=a_4+\sqrt{(q_2/9)^2+(a_4^{(0)})^2} \\[1ex]
 q_5=a_5-(q_2-2)a_3/10  \\[1ex]
 q_6=a_6-\sqrt{(q_2/100)^2+(a_6^{(0)})^2} \;\;.
\end{array} \right.
\label{qi}
\end{equation}
The functions $q_n(\{a_i\})$ were chosen in such a way that the liquid-drop
energy as a function of the elongation $q_2$  becomes minimal along a trajectory
that defines the liquid-drop path to fission. The $a^{(0)}_{2n}$ in Eq.\
(\ref{qi}) are the expansion coefficients of a spherical shape given by
$a^{(0)}_{2n}=(-1)^{n-1}\frac{32}{\pi^3\,(2n-1)^3}$.  The above relations
proposed in Ref.~\cite{SPN17} transform the original deformation parameters
$a_i$ to the more natural parameters $q_i$, which ensure that only minor
variations of the liquid-drop fission paths occur around $q_4=0$. In addition,
more and more elongated prolate shapes correspond to decreasing values of $a_2$,
while oblate ones are described by $a_2>1$, which contradicts the traditional
definition of the elongation parameter. The parametrization (\ref{qi}) is
rapidly convergent. It was shown in Ref.~\cite{Bar20} that the effect of $q_5$
and $q_6$ on the macroscopic potential energy of nuclei is negligible for small
elongations of nuclei up to the saddle points and contributes within 0.5 MeV
around the scission configurations. 

Non-axial shapes can easily be obtained assuming that, for a given value of the
$z$-coordinate, the surface cross-section (blue dashed oval in Fig.~\ref{shape})
has the form of an ellipse with half-axes $a(z)$ and $b(z)$ \cite{SPN17}:
\begin{equation}
   \hspace{-0.3cm}  \varrho_s^2(z,\varphi) = \rho^2_s(z) 
                \frac{1-\eta^2}{1+\eta^2+2\eta\cos(2\varphi)}
   \hspace{0.3cm} \mbox{with} \hspace{0.3cm} 
 \eta = \frac{b-a}{a+b}~,
\label{eta}
\end{equation}
where the parameter $\eta$ describes the non-axial deformation of the nuclear
shapes. The volume conservation condition requires that
$\rho_s^2(z)=a(z)b(z)$.


\subsection{Potential energy surfaces}

The nuclear potential energies of actinide nuclei are evaluated in the following
equidistant grid-points in the 4D collective space built on the $q_2,~q_3,~q_4$,
and $\eta$ deformation parameters:\\[-1ex]
\begin{equation}
\begin{array}{lr}\displaystyle 
q_2=&-0.60~(0.05)~2.35 ~,\\
q_3=&0.00~(0.03)~0.21 ~,\\
q_4=&-0.21~(0.03)~0.21 ~,\\
\eta~=&0.00~(0.03)~0.21 ~.
\end{array}
\label{grid}
\end{equation}

Here, the numbers in the parentheses are the step size, while the numbers on the
left (right) side are the lower (upper) boundaries od the grid, respectively.
The energy of a nucleus is obtained in the mac-mic model, where the smooth
energy part is given by the LSD model \cite{PDu03}, and the microscopic effects
have been evaluated using the YF single-particle potential \cite{DNi76, DPB16}.
The Strutinsky shell-correction method \cite{Str66,NTS69,Pom04} with a
$6^{\mathrm{th}}$ order correctional polynomial and a smoothing width
$\gamma_S=1.2\hbar\omega_0$ is used to determine the shell energy correction,
where $\hbar\omega_0=41/A^{1/3}$ MeV is the distance between the spherical
harmonic-oscillator major shells. The BCS theory \cite{BCS57} with the
approximate GCM+GOA particle number projection method \cite{GPo86} is used for
the pairing correlations. An universal pairing strengths written as $G{\mathcal
N^{2/3}}=0.28\hbar\omega_0$, with ${\mathcal N}=Z, N$ for protons or neutrons,
was adjusted in Ref.~\cite{PPS89} to the experimentally measured mass
differences of nuclei from different mass regions. It was assumed in
Ref.~\cite{PPS89} that the ``pairing window'' contains  $2\sqrt{15{\mathcal N}}$
single-particle energy levels closest to the Fermi level. All the above
parameters were fixed in the past, and none of them was specially fitted to the
properties of actinide nuclei.

\begin{figure}[h!]
\includegraphics[width=0.95\columnwidth]{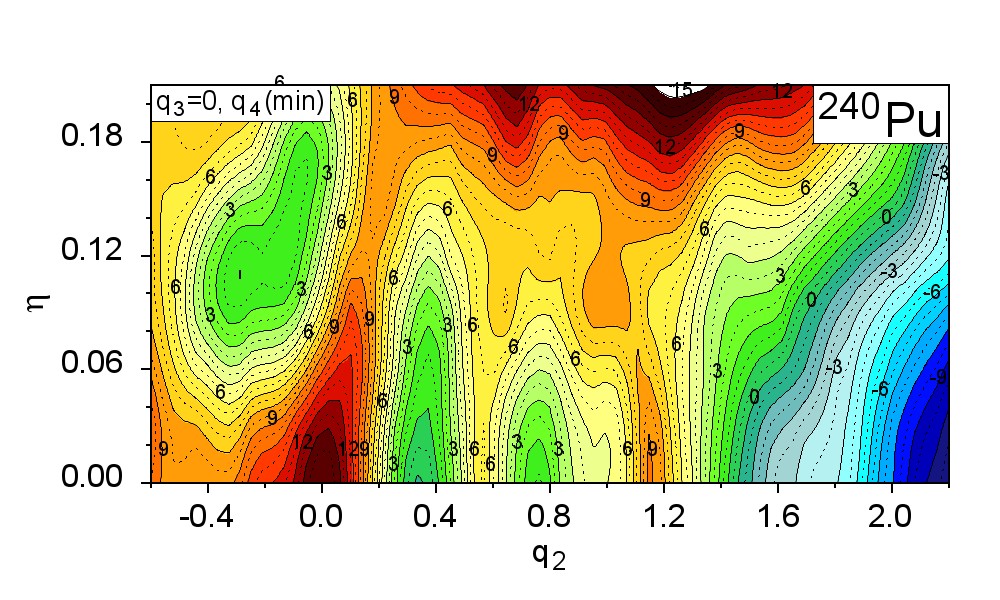}\\[-1ex]
\includegraphics[width=0.95\columnwidth]{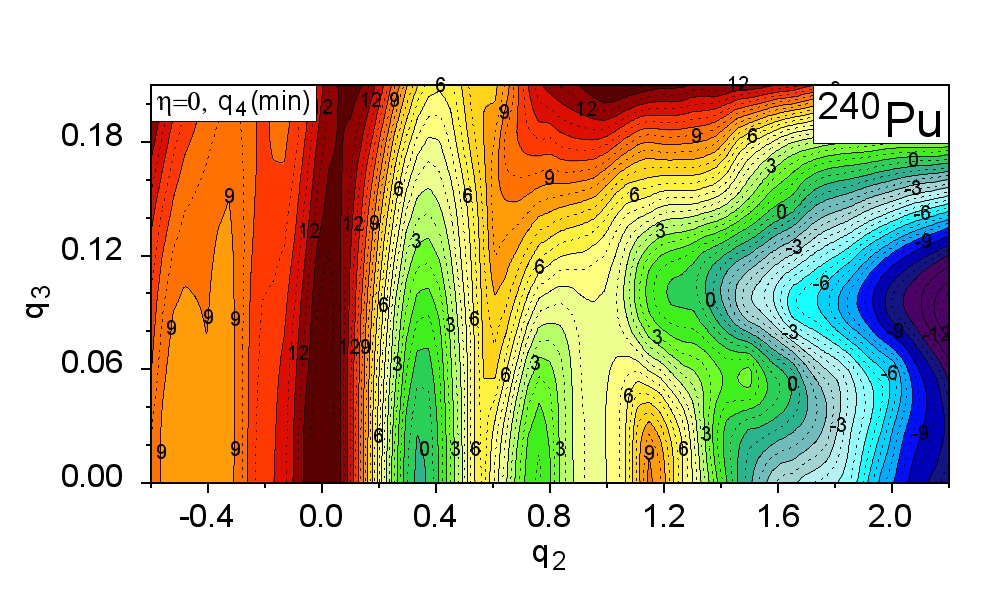}\\[-5ex]
\caption{Potential energy surface of $^{240}$Pu minimized with respect to $q_4$
at the $(q_2,\eta)$ (top map) and $(q_2,q_3)$ (bottom) planes.}
\label{240Pu123}
\end{figure}

A typical PES for actinides is shown in Fig.~\ref{240Pu123}, where two 
cross-sections $(q_2,\eta)$ and $(q_2,q_3)$ of the 4D potential energy surface
of $^{240}$Pu are shown.
As one can see, the inclusion of the non-axial deformation is important up to
elongations corresponding to the second saddle ($q_2\le 1.2$). Apart of some
neutron deficient actinide nuclei which have $q_3\ne 0$ in the ground-state, the
left-right asymmetry  begins to play an important role at large elongations of
nuclei, from the second saddle ($q_2\approx 1$) up to the scission configuration
($q_2 \gtrsim 2$). 
\begin{figure}[h!]
\includegraphics[width=0.95\columnwidth]{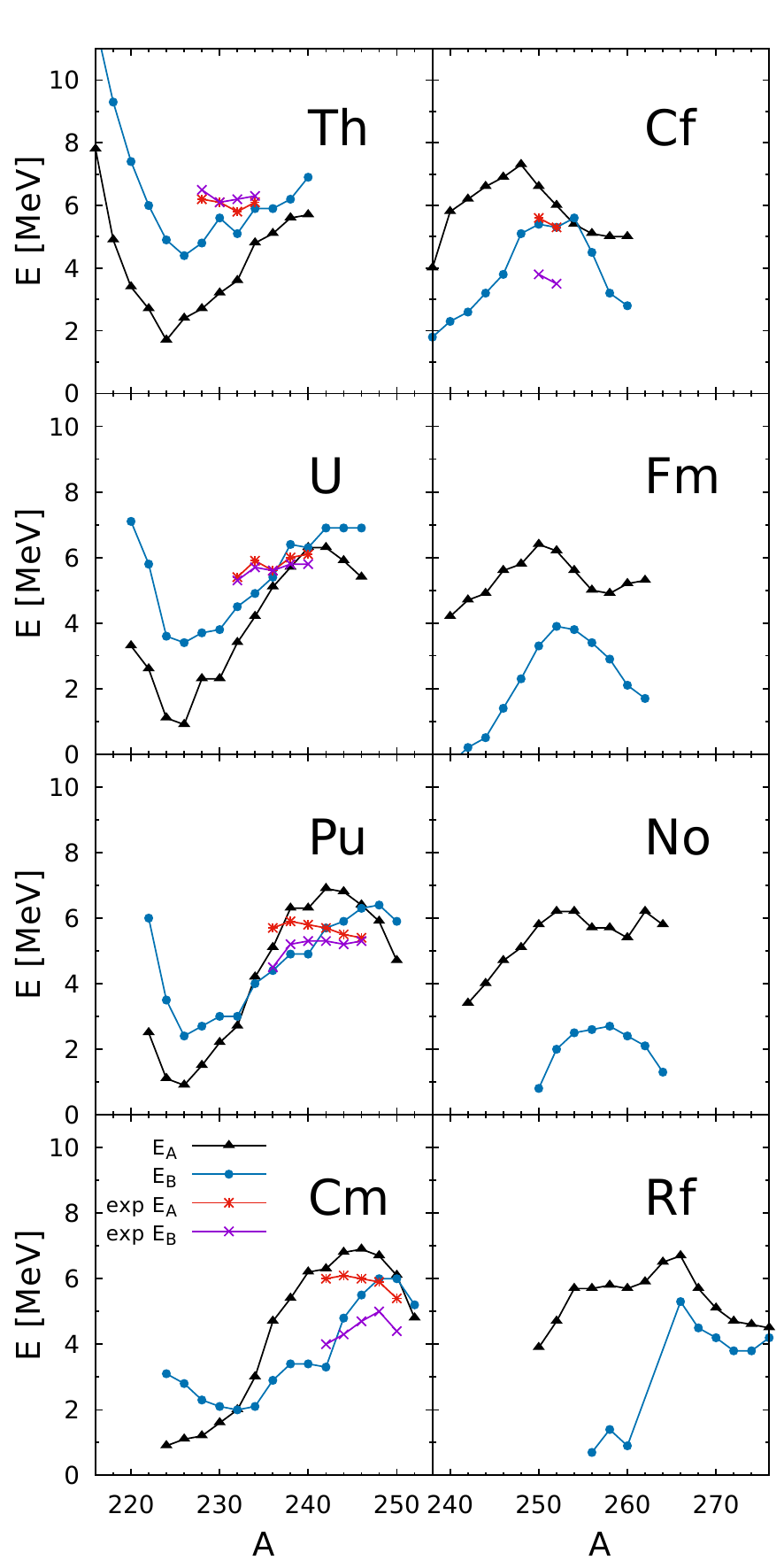}\\[-2ex]
\caption{Fission barrier heights of even-even actinide nuclei in our 4D
  mac-mic model.}
\label{barr}
\end{figure}

The first ($E_A$) and the second ($E_B$) fission barrier heights obtained in 
our model for nuclei from Th to Rf are compared in Fig.~\ref{barr} with the
experimental data taken from Ref.~\cite{Smi93,SJA16}.  The agreement of our
estimates with the data is rather satisfactory, and is comparable within an
accuracy obtained in other theoretical models. The largest deviation between our
estimates and the experimental values are  observed in thorium isotopes, where
they are underestimated. The main origin of these discrepencies is mostly from
the inaccuracy  of determining the ground-state masses in our model. To prove
it, we have  estimated the fission barrier heights using the so called
topographical theorem of Myers  and \'Swi{\c a}tecki \cite{MSw96}, where the
barrier height (the largest one) is defined as 
\begin{equation}
E_{\rm barr}= M_{\rm mac}^{\rm sadd}- M_{\rm g.s.}^{\rm exp}~,
\label{topo}
\end{equation}
where $M_{\rm mac}^{\rm sadd}$ is the first barrier saddle point mass evaluated in the macroscopic model (i.e., without microscopic energy correction) and
$ M_{\rm g.s.}^{\rm exp}$ is the experimental ground-state mass of the nucleus.

\begin{figure}
\includegraphics[width=0.95\columnwidth]{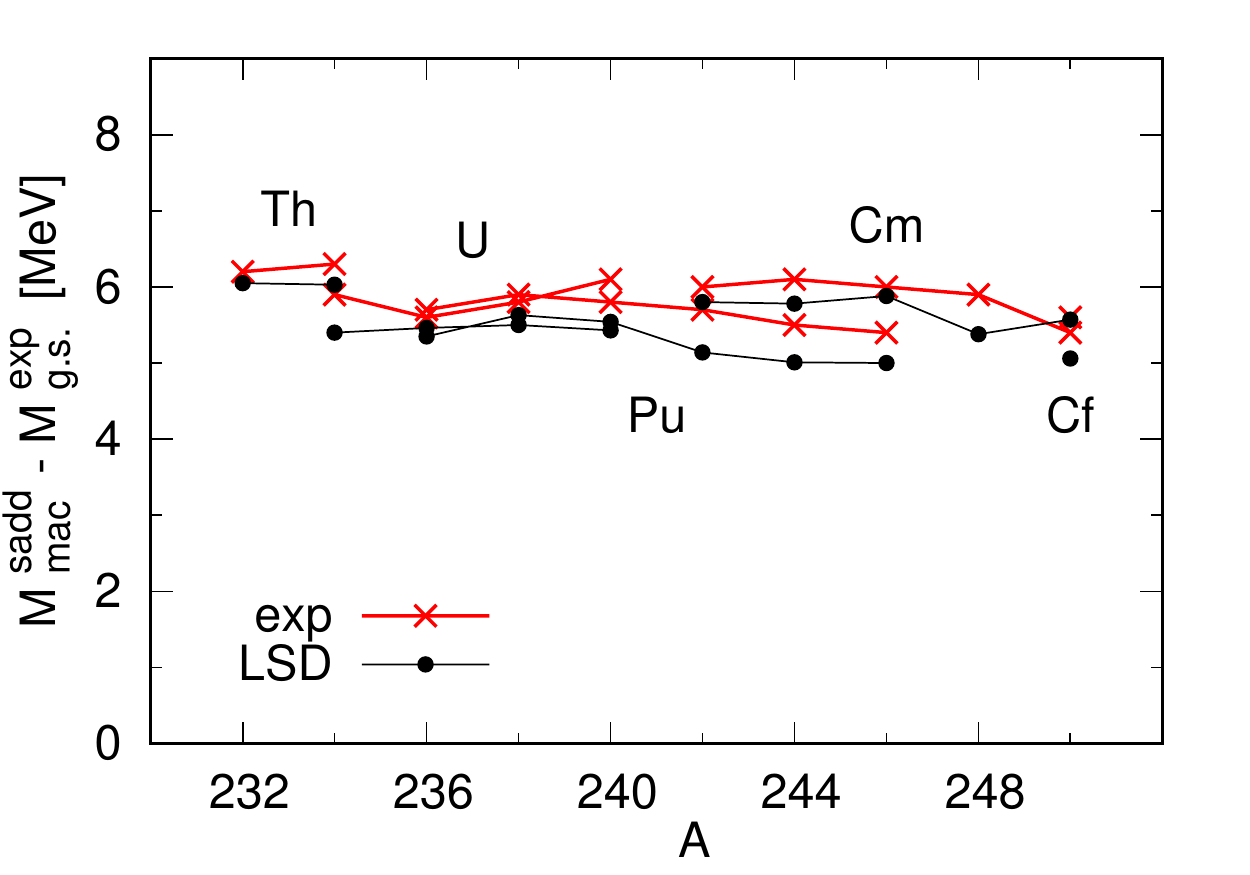}\\[-2ex]
\caption{Fission barrier heights of even-even actinide nuclei evaluated 
using the topographical theorem and the LSD model compared with the
experimental barrier heights as a function of mass number $A$ \cite{DNP09}.}
\label{topobar}
\end{figure}

Using the LSD model \cite{PDu03} to evaluate the macroscopic (read LD) mass one
obtains the '\'Swi{\c a}tecki' estimates of barrier heights, which deviate from
the experimental data only by 310 keV on the average, as shown in Fig.~\ref{topobar}.
It means that additional work to improve the estimates of the ground-state
masses has to be done. In particular, to make a better fit of the pairing
strength. Our ''universal'' pairing force \cite{PPS89}, used in the present
work, reproduces on average the pairing gaps of nuclei from different mass
regions, but it might be that it does not reproduce perfectly the pairing
properties in actinides.

\subsection{Simple collective model for fission}

The present research is a continuation and extension of our previous works
\cite{PIN17,PNB18,PDH20}, where more detailed description of the collective
fission model was given. The fundamental idea of this approach is the use of the
Born-Oppenheimer approximation (BOA) to separate the relatively slow motion
towards fission, mainly in  $q_2$ direction, from the fast vibrations in the
``perpendicular'' $q_3$ and $q_4$ collective coordinates. The BOA allows us to
treat these both types of motion as decoupled, what leads, in consequence, to
the wave function in form of the following product:\\[-3ex]
\begin{equation}
 \Psi_{nE}(q_2,q_3,q_4)=u_{nE}(q_2)\,\phi_n(q_3,q_4;q_2)~.
\end{equation}
The function $u_{nE}(q_2)$ is the eigenfunction corresponding to the motion
towards fission, while the $\phi_n(q_3,q_4;q_2)$ simulates the $n-$phonon
``fast'' collective vibrations in the ``perpendicular'' to the fission mode
$\{q_3,q_4\}$ plane.

To determine the $u_{nE}(q_2)$ function for a single $q_2$ mode one can use the WKB approximation as it has been done in Ref.~\cite{PIN17}. To obtain the
function $\phi_n(q_3,q_4;q_2)$, one has to solve numerically for each value of
$q_2$ the eigenproblem of the underlying Hamiltonian in the perpendicular
$\{q_3,q_4\}$ space. However, for the low energy fission, it is sufficient
to take only the lowest wave function in the perpendicular mode and evaluate 
the density of probability $W(q_3,q_4;q_2)$ of finding the system for a given 
elongation $q_2$ within the area of $(q_3\pm dq_3, q_4\pm dq_4)$ as 
\begin{equation}
W(q_3,q_4;q_2)= |\Psi(q_2,q_3,q_4)|^2=|\phi_0(q_3,q_4;q_2)|^2~.
\label{W_prob}
\end{equation}
Further simplification we have made is to approximate the modulus square of the total wave function in Eq.~(\ref{W_prob}) by the Wigner function in the 
following form
\begin{equation}
W(q_3,q_4;q_2)\propto \exp{\frac{V(q_3,q_4;q_2)-V_{\rm min}(q_2)}{T^*}}~,
\label{W_prob1}
\end{equation}
where $V_{\rm min}(q_2)$ is the minimum of the potential for a given elongation
$q_2$ and $T^*$ is a generalized temperature \cite{WWH81} which takes into 
account both thermal excitation of fissioning nucleus and the collective 
zero-point energy $E_0$
\begin{equation}
T^*= E_0/{\rm tanh}(E_0/T)~.
\label{tstar}
\end{equation}
The temperature ($T$) of nucleus with mass-number $A$ is evaluated from its
thermal excitation energy ($E^*$) using the phenomenological relation
$E^*=aT^2$, with $a=A$/(10 MeV).  The generalized temperature $T^*$  is
approximately equal to the zero-point energy when $T$ is small while for
sufficiently high temperatures ($T\gg E_0$) it approaches to $T$. In the
following $E_0$ is treated as one of two adjustable parameters of our model. Of
course, one expects $E_0$ of the order of 1 to 2 MeV as implied by the energy
level positions of typical collective vibrational states.

To obtain the FMY for a given elongation $q_2$ one has to integrate the 
probabilities (\ref{W_prob1}) over the full range of the neck parameter
$q_4$\\[-3ex]
\begin{equation}
w(q_3;q_2)=\int W(q_3,q_4;q_2) dq_4~.
\label{W_integr}
\end{equation}
It is rather obvious that the fission probability may strongly depend on the
neck radius $R_{\rm neck}$. Following Ref.~\cite{PIN17} one assumes the neck
rupture probability $P$ to be equal to
\begin{equation}
P(q_2,q_3,q_4)=\frac{k_0}{k}\,P_{\rm neck}(R_{\rm neck})~,
\label{neck_rap}
\end{equation}
where $P_{\rm neck}$ is a geometrical factor indicating the neck breaking
probability proportional to the neck thickness, while $k_0/k$ describes the fact
that the larger collective velocity towards fission, $v(q_2)=\dot q_2$, implies
that the neck rupture between two neighboring $q_2$ configurations is getting
less probable. The  constant parameter $k_0$ plays the role of scaling parameter
which is finally eliminated in the calculation of the resulting FMY. The
expression for the geometrical probability factor $P_{\rm neck}(R_{\rm neck})$
is chosen here in a form of Gauss function \cite{PNB18}:
\begin{equation}
P_{\rm neck}(R_{\rm neck})=\exp{[-\log(2)(R_{\rm neck}/d)^2]}~,
\label{P_neck}
\end{equation}
where $d$, our second adjustable parameter, is the ``half-width'' of the neck-breaking probability. The momentum $k$ in Eq.~(\ref{neck_rap}) simulates
the dynamics of the fission process, which, as usual, depends both on the local
collective kinetic energy ($E_{\rm kin}$) and the inertia ($M$) towards the 
fission mode
\begin{equation}
\frac{\hbar^2 k^2}{2\bar{M}(q_2)}=E_{\rm kin}=E-E^*-V(q_2) ~,
\label{energy_cons}
\end{equation}
with $\bar{M}(q_2)$ standing for the (averaged over $q_3$ and  $q_4$ degrees of
freedom) inertia parameter at a given elongation $q_2$, and $V(q_2)$ is the
potential corresponding to the bottom of the fission valley. In the further
calculations we assume that the part of the total energy converted into heat
$E^*$ is negligibly small due to rather small friction forces in the low energy
fission. A good approximation of the inertia $\bar{M}(q_2)$, proposed in
Ref.~\cite{RLM76}, is to use the irrotational flow mass parameter $B_{\rm irr}$,
which is derived initially as a function of the distance between fragments
$R_{12}$ and the reduced mass $\mu$ of both fragments
\begin{equation}
\bar{M}(q_2)=\mu [1+11.5\,(B_{\rm irr}/\mu -1)]\bigg(\frac{\partial R_{12}}
{\partial q_2}\bigg)^2 ~.
\label{irrM}
\end{equation}
In order to make use of the neck rupture probability $P(q_3,q_4;q_2)$ of
Eq.~(\ref{neck_rap}), one has to rewrite the integral over $q_4$ in probability
distribution (\ref{W_integr}) in the following form:
\begin{equation}
w(q_3;q_2)=\int W(q_3,q_4;q_2) P(q_2,q_3,q_4) dq_4 ~,
\label{W_integrP}
\end{equation}
in which the neck rupture probability is now taken into account.
The above approximation describes a very important fact that, for a fixed $q_3$
value, the fission may occur within a certain range of $q_2$ deformations with
different probabilities. Therefore, to obtain the true fission probability
distribution $w'(q_3;q_2)$ at a strictly given $q_2$, one has to exclude the
fission events occurred in the ``previous'' $q'_{2}<q_2$ configurations, i.e.,
\begin{equation}
w'(q_3;q_2)=w(q_3;q_2)\frac{1-\int\limits_{q'_{2}<q_2} w(q_3;q'_2) dq'_2}
{\int w(q_3;q'_2) dq'_2}~.
\label{w_prev}
\end{equation}

The normalized mass yield is then obtained as the integral of partial yields
over $q_2$:
\begin{equation}
Y(q_3)=\frac{\int w'(q_3;q_2) dq_2}{\int w'(q_3;q_2) dq_3 \, dq_2}~.
\label{Yield}
\end{equation}

Since there is a one-to-one correspondence between $q_3$ deformation and the
masses of the left ($A_L$) and right ($A_R=A-A_L$) fission fragments, the yield given by Eq.~(\ref{Yield}) can be directly compared with the experimental FMY's.
Note that due to the normalization procedure (\ref{Yield}), the scaling 
parameter $k_0$ introduced in Eq.~(\ref{neck_rap}) does no longer appear in 
the definition of mass yield. 

So, there are only two free parameters in the above model, namely the zero-point
energy $E_0$ in Eq.~(\ref{tstar}) and the half-width parameter $d$ appearing in
the probability of neck rupture in Eq.~(\ref{P_neck}). 

\section{Fission fragment mass-yields}

Relatively good estimates of the FMY's obtained in our previous works
\cite{PNB18,PDH20} for Pu and Pt to Ra isotopes encourage us to apply our model
to describe and predict the mass yield for the low-energy fission of actinide
nuclei from Th to Rf.  Our main  goal is to show that the innovatory
Fourier shape parametrization \cite{PNB15} and the mac-mic model based on the
LSD macroscopic energy \cite{PDu03} with  the microscopic energy correction
evaluated using the Yukawa-folded potential \cite{DNi76,DPB16} describes well
the existing fission valleys in the broad region of nuclei. It is worth
recalling that none of the model parameters, apart from $E_0$ and $d$ described in the previous section, was modified here to get a better description. 

In general, the theoretical estimates of FMY depend weakly on the choice of the
free parameters $E_0$ and $d$. To obtain the best fit to the existing
experimental FMY's, the following overlap of  theoretical and experimental
yields:
\begin{equation} 
I(E_0,d)=\sum\int\limits_0^A |Y_{exp}(A_f)-Y_{th}(A_f;E_0,d)| dA_f ~.
\label{fit}
\end{equation}
is minimized with respect to $E_0$ and $d$. The sum in Eq. (\ref{fit}) runs over
the nuclei, where the experimental data exist. Contrary to the $\xi^2$ fit, this
fitting procedure does not overestimate the role of large deviations. The
optimised values $E_0=2.2$ MeV and $d=$1.6 fm are finally used to obtain the FMY
for all considered actinide nuclei.

It is well known that the FMY of a given nucleus is mainly determined by its PES
properties at large deformations. Typical examples of PES for $^{240}$Pu are
shown in Fig.~\ref{240Pu123}, where the mac-mic energy minimized with respect to
$q_4$ is plotted on the $(q_2,\eta)$ (top) and $(q_2,q_3)$ (bottom) planes. The
labels at the layers correspond to the energy of the deformed nucleus (in MeV)
measured with respect to the LSD macroscopic energy of the spherical nucleus.
The first saddle is visible around $q_2=0.55$ and $q_3=0$, while the second one
is at $q_2=1.10$ and $q_3=0.08$. As one can see in the upper panel, the
non-axial deformation $\eta$ does not influence the PES at lager $q_2$
deformation. So, we do not take this degree of freedom into account in our
analysis of the  FMY's. Let us notice that each of the 2D energy maps shown in
Fig. \ref{240Pu123} is only a projection of the full 4D PES, and one has to
consider other cross-sections in order to analyze the fission process in
details. 

\begin{figure}[htb]
\includegraphics[width=0.95\columnwidth]{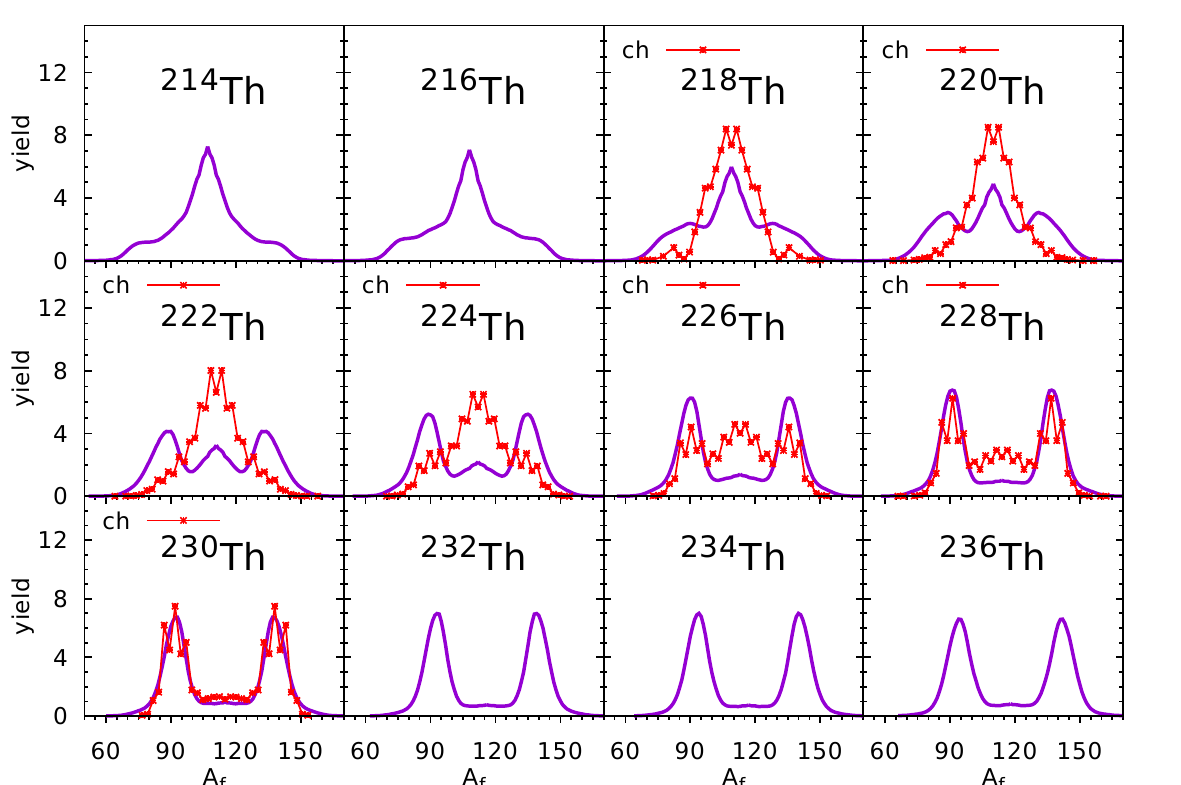}
\includegraphics[width=0.95\columnwidth]{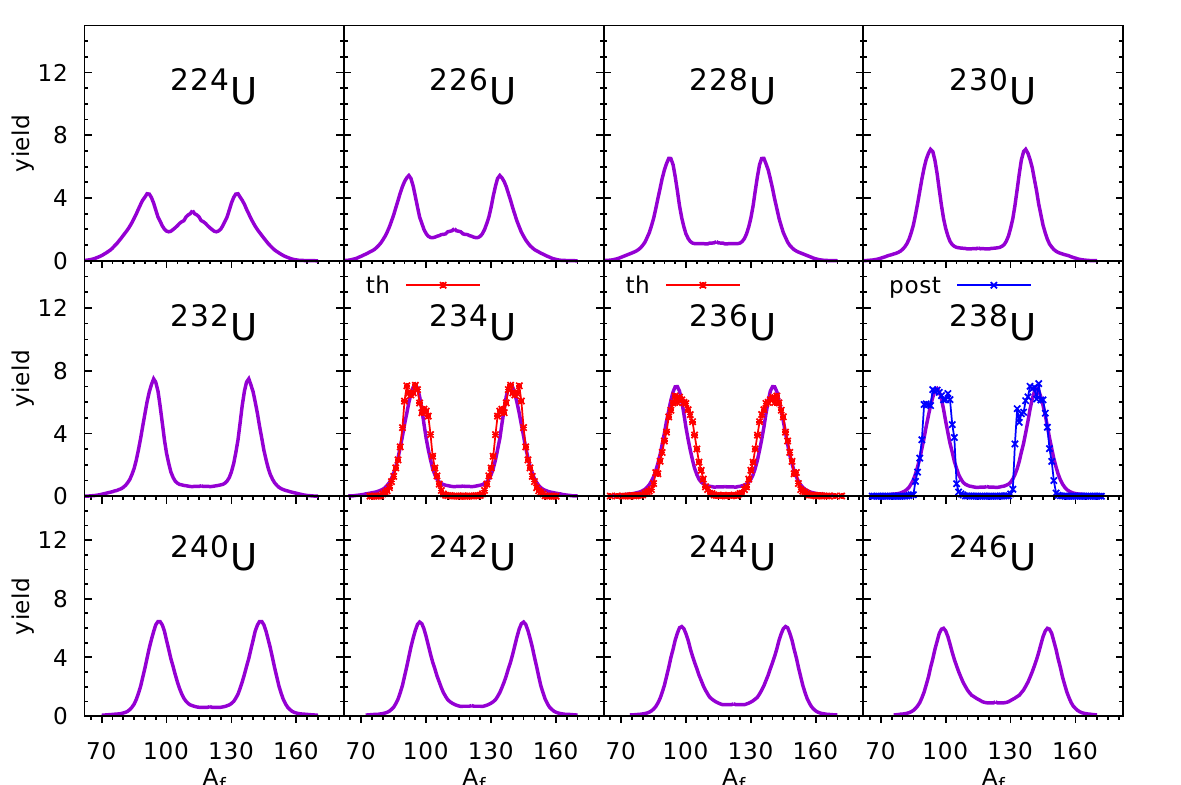}\\[-2ex]
\caption{Fission fragment mass-yields of Th (top part) and U (bottom part)
isotopes. Experimental data (red stars) for Th isotopes are extracted from the
charge-yields of Refs.~\cite{Sch00,Chat19} while the mass-yields for U isotopes
(botton part) are taken from Ref.~\cite{Gel85,Rom10} for the thermal neutron 
induced fission (th). Just to guide the eye we have used for $^{238}$U the 
post-neutron data (blue crosses) taken from Ref.~\cite{SJA16}.}
\label{Th-U}
\end{figure}
\begin{figure}[htb]
\includegraphics[width=0.95\columnwidth]{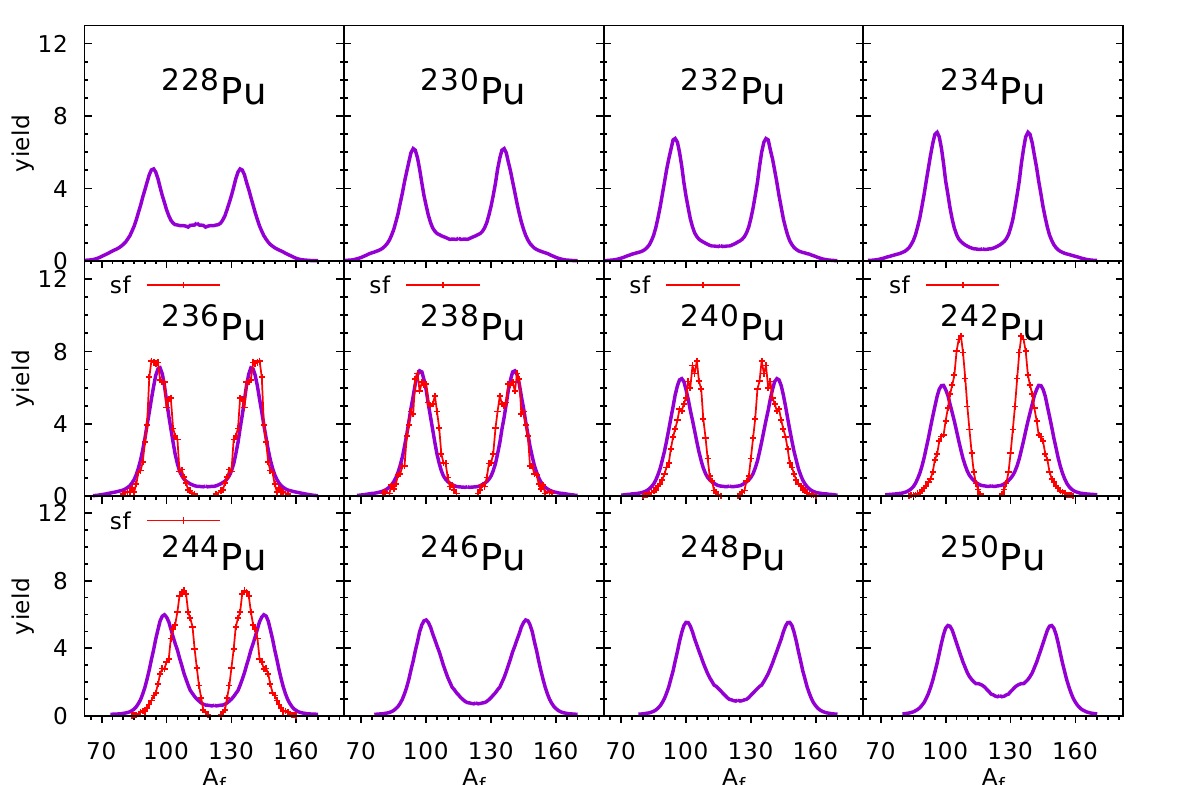}
\includegraphics[width=0.95\columnwidth]{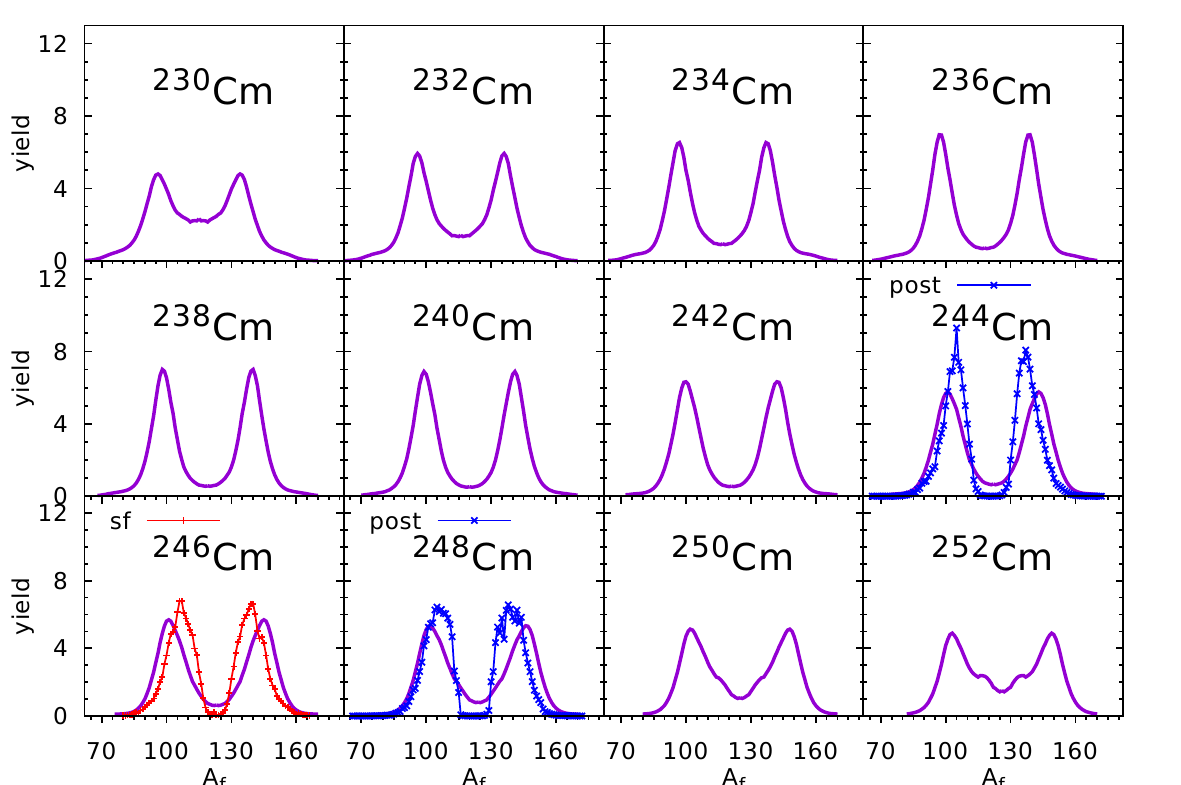}\\[-2ex]
\caption{Fission fragment mass yields of Pu (top part) and Cm (bottom part)
         isotopes. Experimental data (red stars) are taken from 
         Ref.~\cite{Dem97} for Pu chain and Refs.~\cite{SJA16,Ple73} for Cm
         nuclei.}
\label{Pu-Cm}
\end{figure}
\begin{figure}[htb]
\includegraphics[width=0.95\columnwidth]{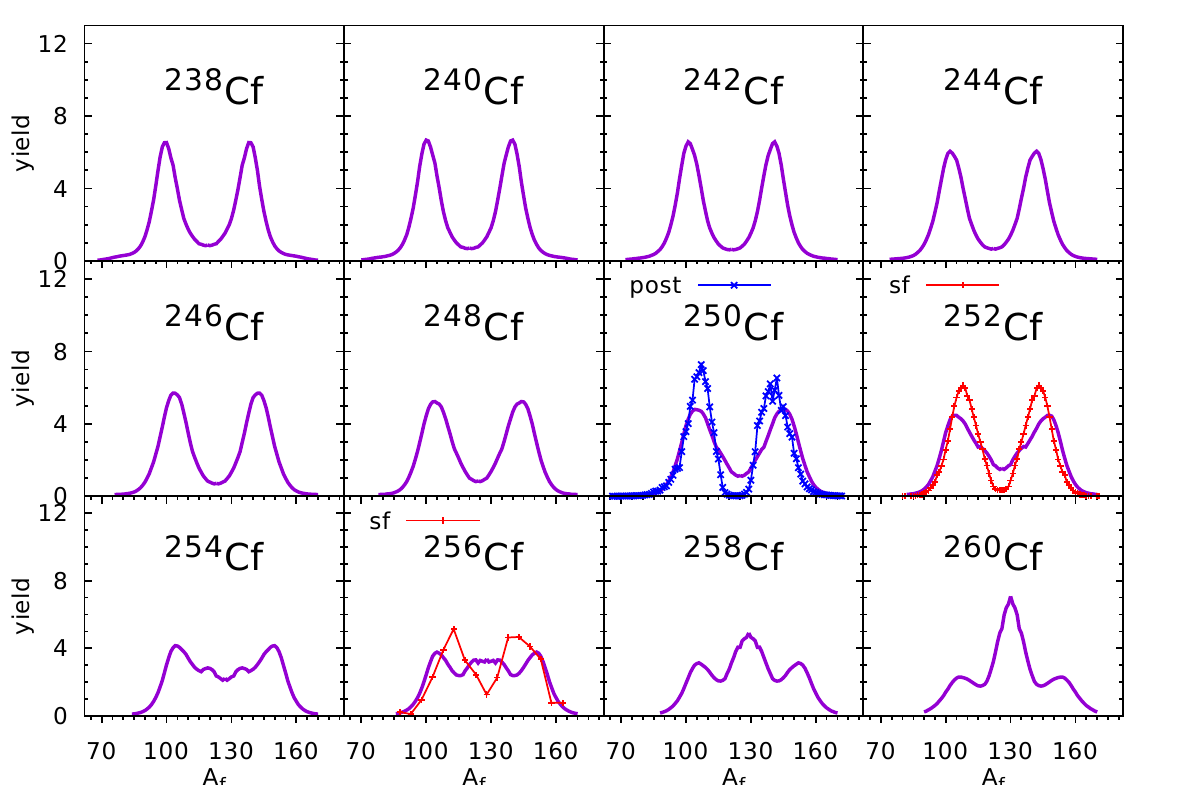}
\includegraphics[width=0.95\columnwidth]{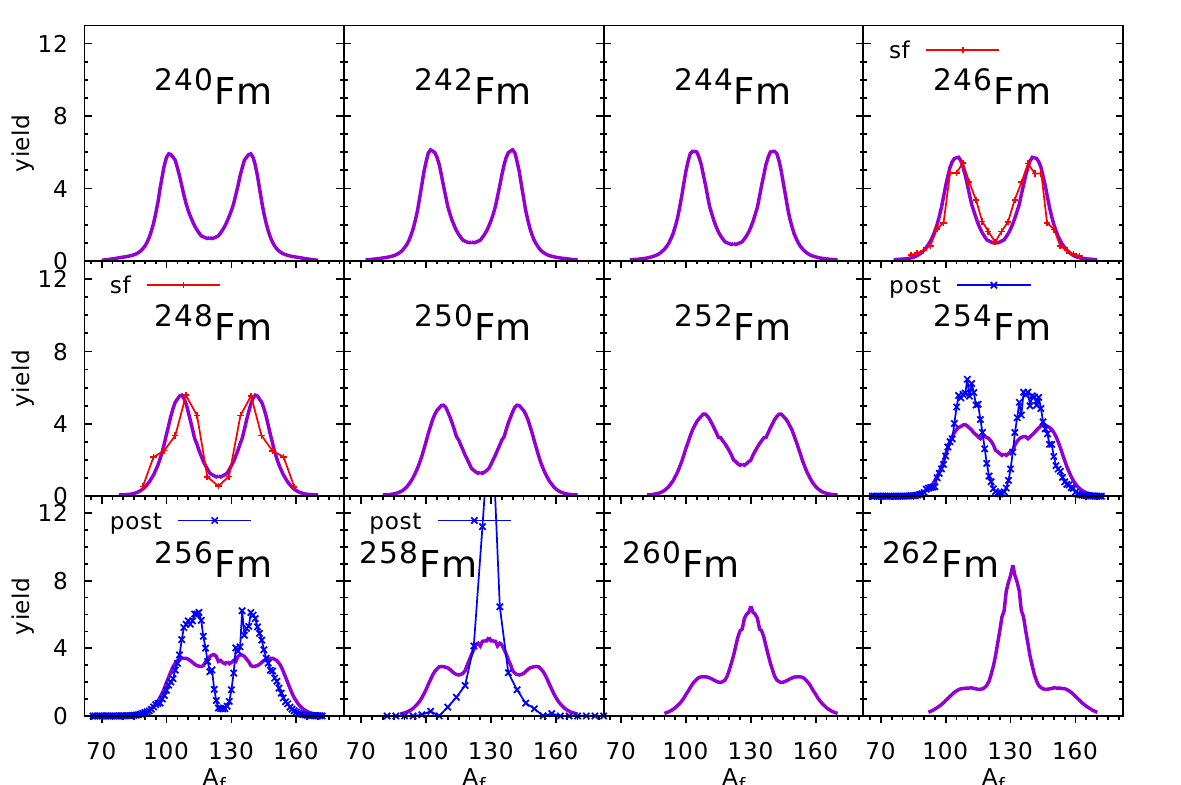}\\[-2ex]
\caption{Fission fragment mass yields of Cf (top) and Fm (bottom) isotopes. 
Experimental data for pre-neutron yields (red stars) are taken from Refs.
\cite{Rom10,Hof80} while the post-neutron yields (blue crosses) origin from 
Ref.~\cite{SJA16,Hul86}.}
\label{Cf-Fm}
\end{figure}
\begin{figure}[t!]
\includegraphics[width=0.95\columnwidth]{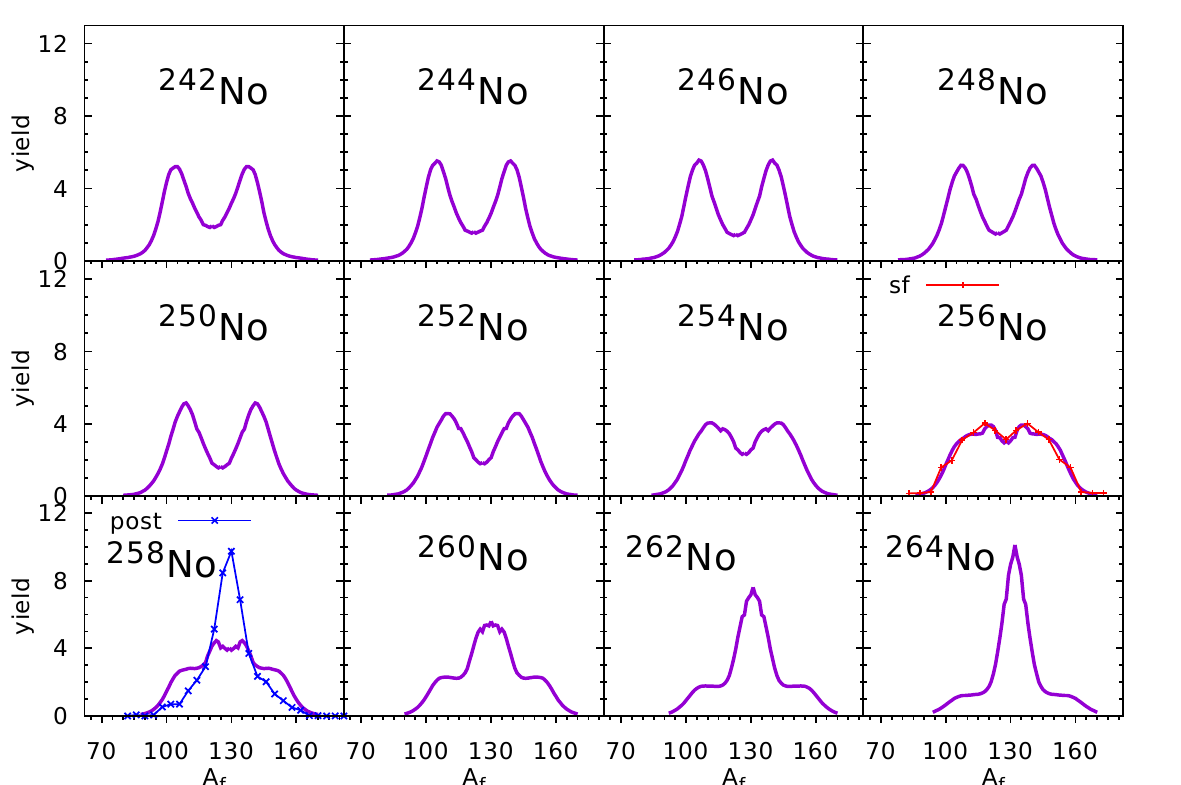}
\includegraphics[width=0.95\columnwidth]{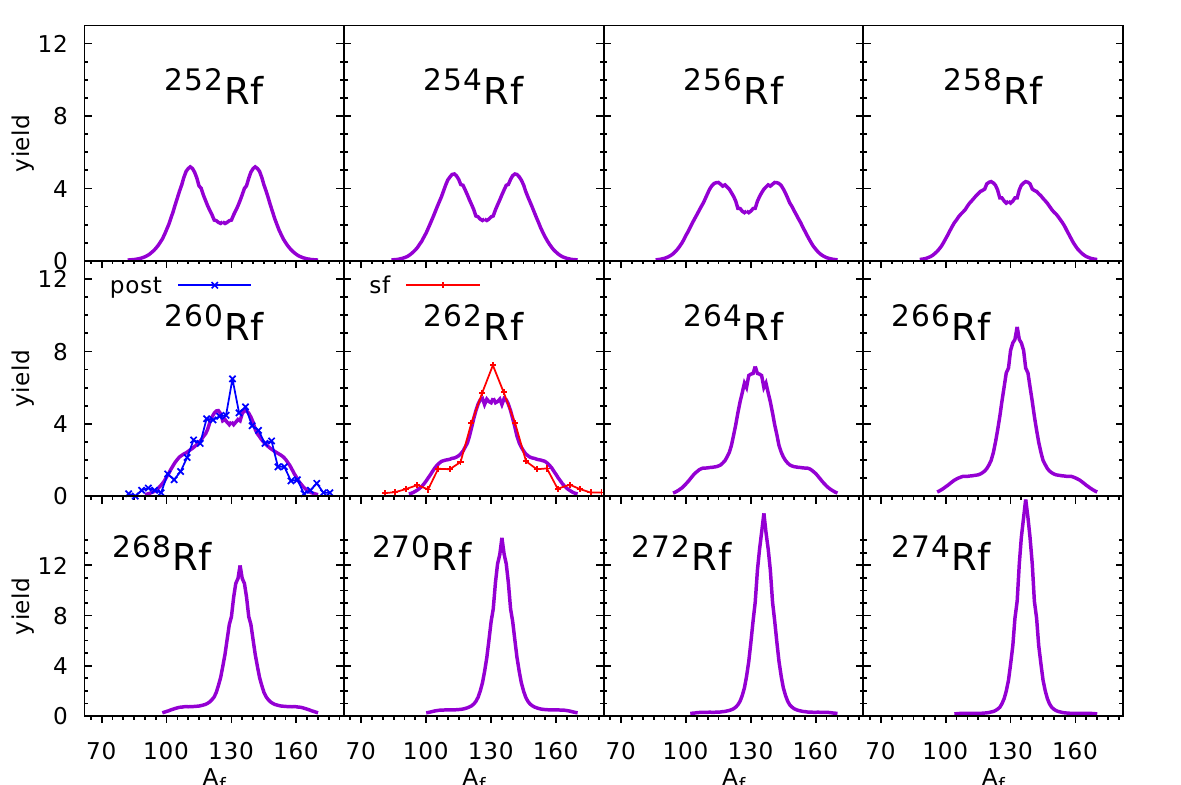}\\[-2ex]
\caption{Fission fragment mass yields of No (top) and Rf (bottom) isotopes.
Experimental data (crosses) are taken from Refs.~
\cite{Hof90,Hul86,Lan96}.}
\label{No-Rf}
\end{figure}

The fission fragment mass yields obtained in our model are presented in Figs.
\ref{Th-U} to \ref{No-Rf}. Some experimental data for the FMY were obtained for
the fission of excited nuclei. In such a case, we take this excitation into
account and reduce the microscopic energy correction according to the
prescription found in Ref.~\cite{NPB06}. Our estimates of FMY correspond to  the
so-called pre-neutron yields, i.e., the mass yields before neutron emission from
fragments and with such data (red stars in Figs.~\ref{Th-U} to \ref{No-Rf})
they have to be compared. In the case of Th isotopes, we have used the fragment
charge yields from Refs. \cite{Sch00,Chat19} and to obtain the mass-yields it is
assumed that the $Z/N$ ratio in the fragment is the same as in the mother
nucleus. In cases when the pre-neutron data were not available, we have plotted
the post-neutron data (blue crosses) just to get piece of information about the
experimental situation. It is shown that for the Th isotopes, although the
agreement of the estimates with the experimental data is not very satisfactory,
the general trend is reproduced, i.e., a transition from symmetric to asymmetric
fission is evidently reproduced with a growing mass number of isotope. The best
agreement was achieved for $^{218}$Th and $^{228-230}$Th nuclei. The agreement
with experimental data in the Uranium chain, presented in the bottom part of
Fig.~\ref{Th-U}, is much better. Here the maxima and the widths of the fragment
mass distribution are well reproduced, and a similar transition between
symmetric and asymmetric fission as in Th isotopes is evident.  

The prediction of the  FMY's for Pu and Cm isotopes are compared in
Fig.~\ref{Pu-Cm} with the experimental data. The pre-neutron experimental yields
for Pu \cite{Dem97} and $^{246}$Cm \cite{Ple73} isotopes are obtained for the
spontaneous fission case, while those for $^{244}$Cm and $^{248}$Cm are
post-neutron yields taken from Ref. \cite{SJA16}. A nice agreement with the data
obtained for the two lightest Pu isotopes $^{236}$Pu and $^{238}$Pu  is slightly
spoiled when the number of neutrons  increases i.e. for $^{242}$Pu and
$^{244}$Pu. It is mainly because we have used here the globally optimized values
of $E_0$ and $d$, which are not fitted to Pu data only as done in
Ref.~\cite{PNB18}. In all investigated Pu and Cm isotopes here, the asymmetric
fission is predicted with the mass of the heavy fragment $A\approx 140$. 

The estimates of the FMY for Cf and Fm chains of isotopes are shown in Fig.
\ref{Cf-Fm}. All experimental data correspond to the spontaneous fission, but
only for $^{252}$Cf \cite{Rom10} and $^{256}$Cf \cite{Hof80} have to do with
pre-neutron yields. The rest of the experimental yields presented in 
Fig.~\ref{Cf-Fm} corresponds to the post-neutron data (blue crosses). One can
see that for lighter Cf and Fm isotopes, the asymmetric yields are predicted,
while in the case of the heaviest Cf and Fm nuclei, the symmetric fission is
foreseen. As one can deduce from the above results, our estimates are rather
consistent with the experimental yields.
\begin{figure}[h!]
\includegraphics[width=0.95\columnwidth]{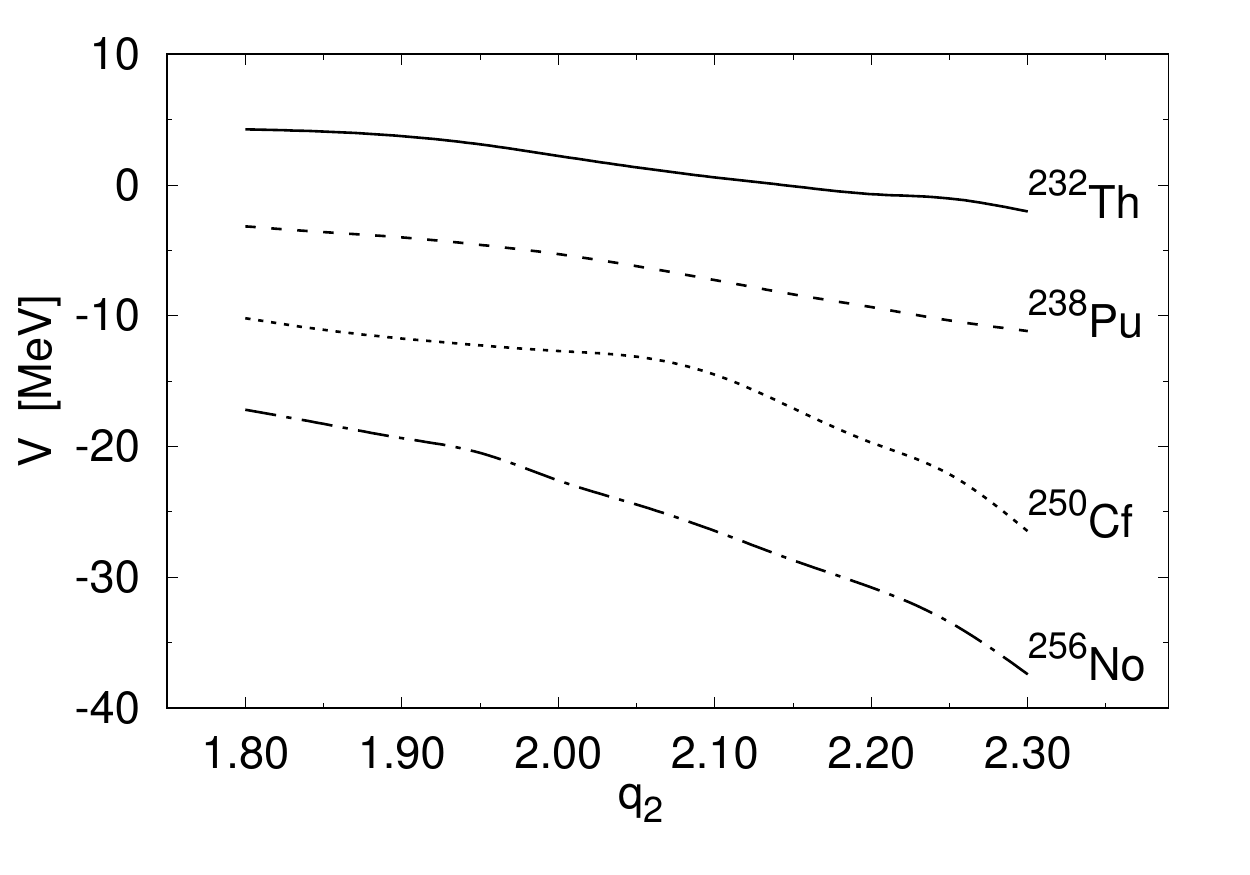}\\[-2ex]
\caption{Potential corresponding to the bottom of fission valley as a function
of the elongation parameter $q_2$}  
\label{slope}
\end{figure}
A similar tendency as seen in the Cf and Fm chains can be observed in Fig.
\ref{No-Rf} for the No and Rf isotopes, where the asymmetric fission is 
predicted in the lighter nuclei while the symmetric fission mode dominates for
the isotopes with $N > 156$. The agreement with the experimental data of
$^{256}$No and $^{262}$Rf \cite{Hof90,Lan96} is evident. 

The overall good quality of our predictions in a broad mass region of the
actinide elements is probably due to the fact that in very heavy nuclei, the
fission barrier is very short, and the fission valley forms very early, i.e., at
a relatively small elongation of the nucleus. An oposite situation occurs in the
thorium nuclei, where the fission barriers are very broad.  Fig.~\ref{slope}
presents the fission valley potential as a function of the elongation parameter
$q_2$. It is shown that the average slope of the curve from the last saddle to
scission in the thorium nuclei is almost three times smaller than in nobelium.
Obviously, such a large difference in the slope towards fission influences the
fission dynamics in these both types of nuclei. This is a main reason why one
has to study in details the PES in Th nuclei to explain here observed change in
the FMY systematics.
\begin{figure}[t!]
\includegraphics[width=0.95\columnwidth]{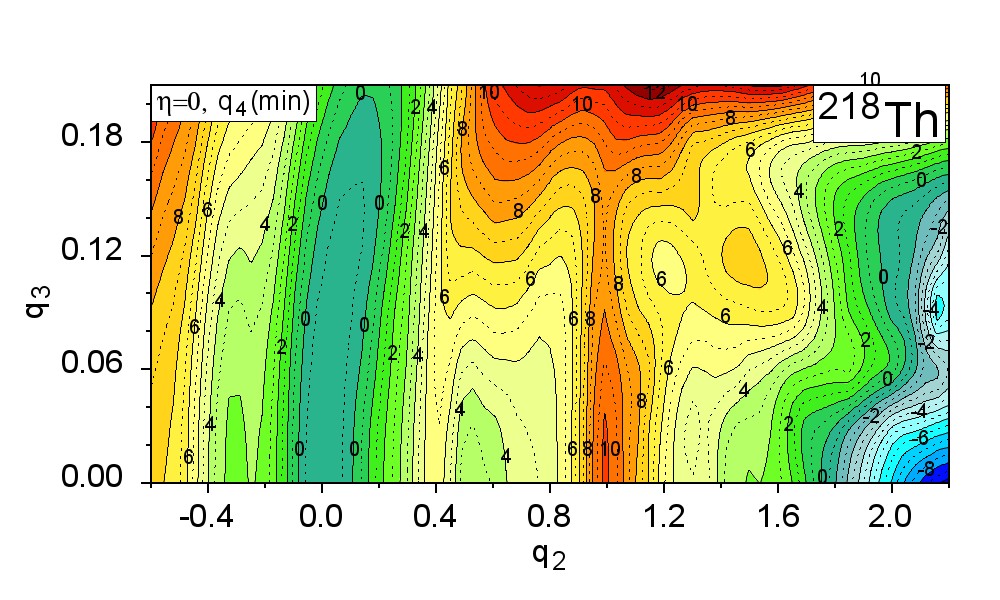}\\[-2ex]
\includegraphics[width=0.95\columnwidth]{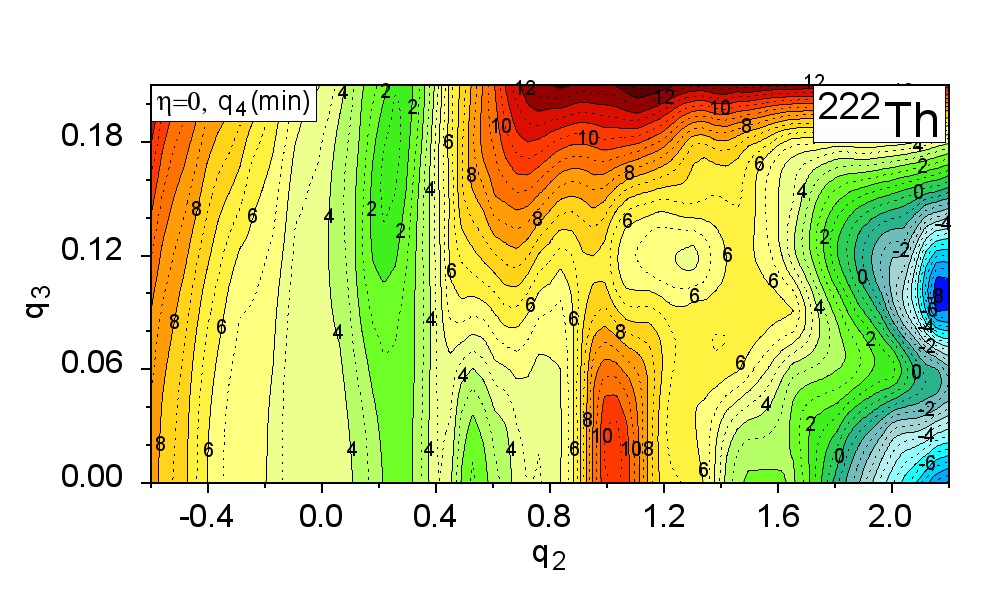}\\[-2ex]
\includegraphics[width=0.95\columnwidth]{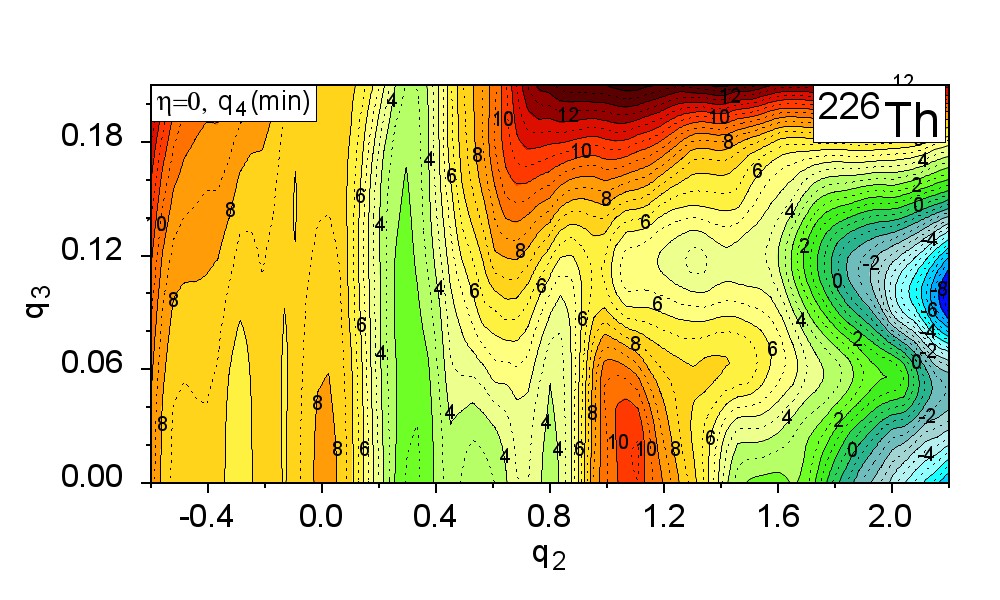}\\[-2ex]
\includegraphics[width=0.95\columnwidth]{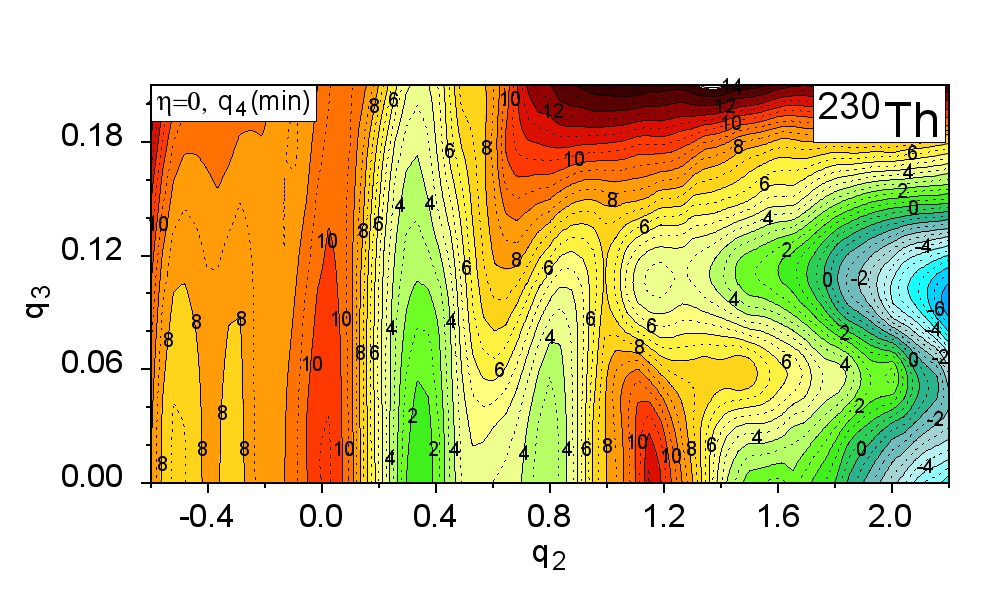}\\[-2ex]
\caption{Potential energy surface cross-sections of $^{218-230}$Th isotopes
minimized with respect the neck parameter $q_4$ on the plane $(q_2,q_3)$.} 
\label{Th23}
\end{figure}

\section{Specific discussions on Th nuclei}

The agreement of our estimates of the FMY's with the experimental data in the Th
chain of isotopes depicted in Fig.~\ref{Th-U} is not quantitatively
satisfactory. So, in the present section, we would like to look for the origin
of these discrepancies. First, these yields for the Th nuclei are evaluated
using $E_0$ and $d$ obtained by the fit to the data for all nuclei. The PES's
for Th nuclei are very much different from those for heavier nuclei. This can be
seen by comparing the PES of $^{240}$Pu shown in Fig.~\ref{240Pu123} (bottom)
with the corresponding maps for $^{218-230}$Th isotopes presented in
Fig.~\ref{Th23}. 
\begin{figure}[b!]
\includegraphics[width=0.95\columnwidth]{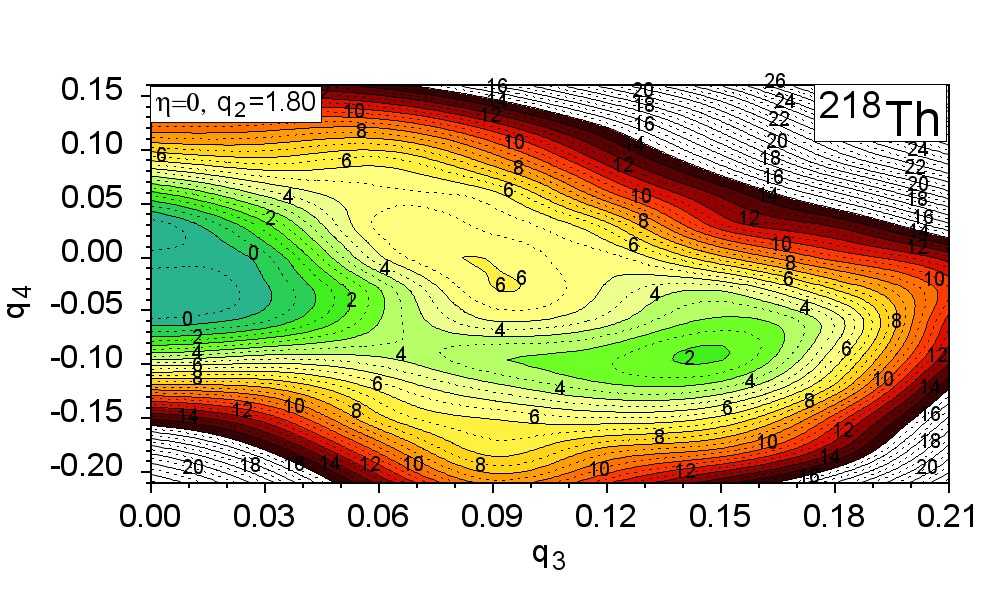}\\[-2ex]
\includegraphics[width=0.95\columnwidth]{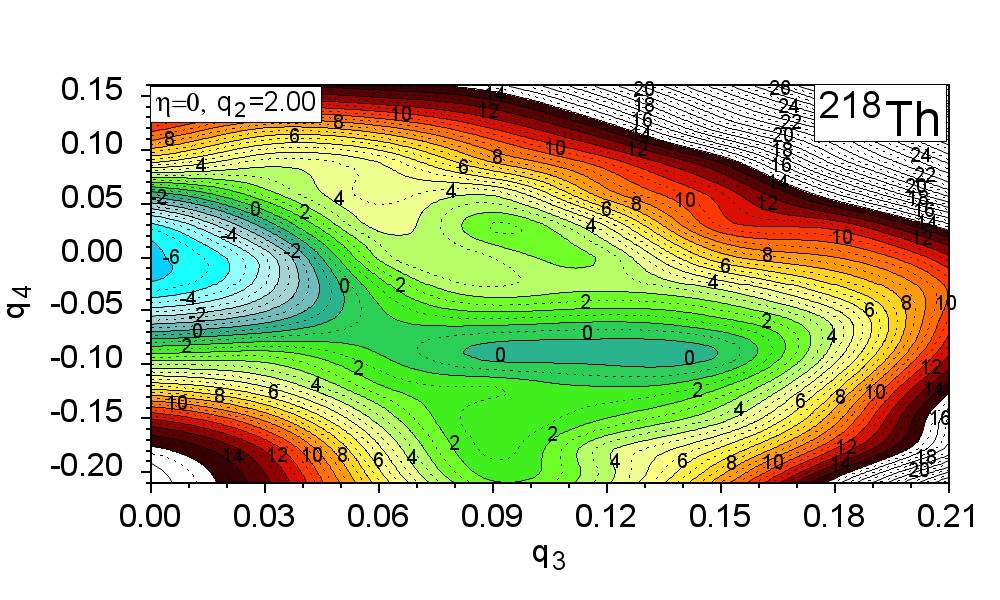}\\[-2ex]
\includegraphics[width=0.95\columnwidth]{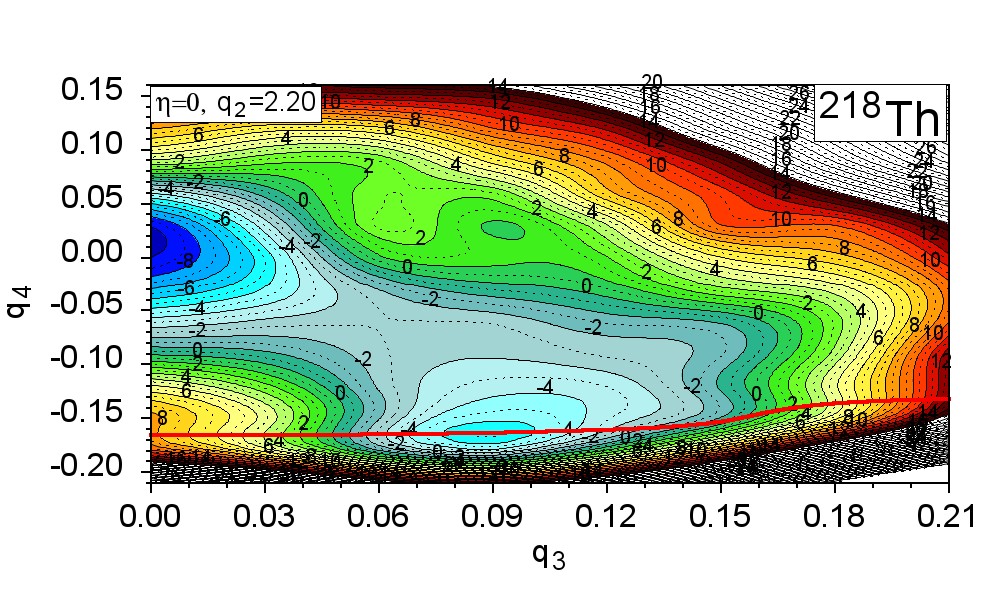}\\[-2ex]
\includegraphics[width=0.95\columnwidth]{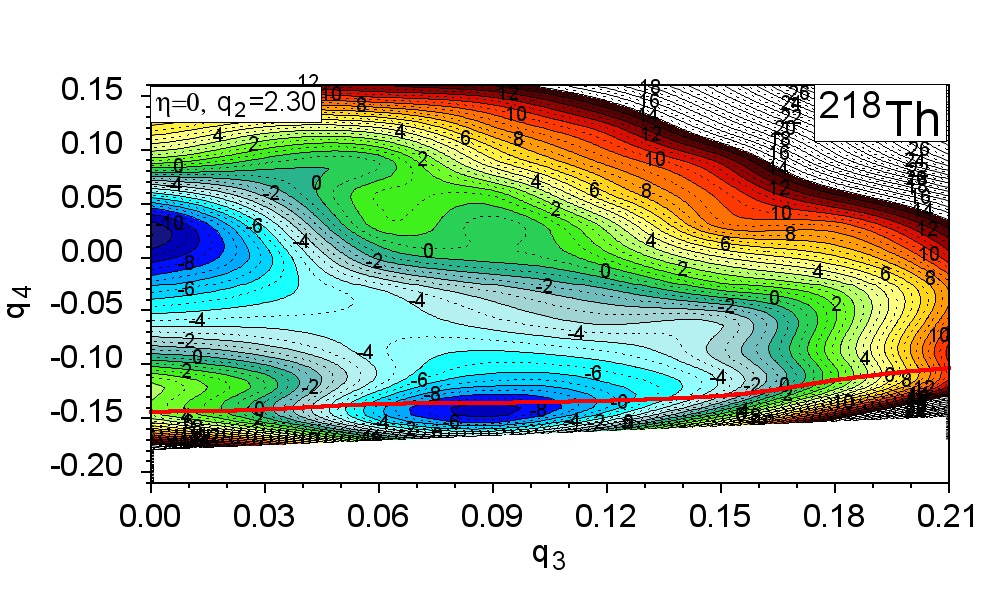}\\[-2ex]
\caption{Potential energy surface cross-sections of $^{218}$Th on the plane
$(q_3,q_4)$. The panels from top to bottom correspond to elongations $q_2=$ 1.8
to 2.3, respectively. The solid red lines drawn in the bottom  panels
correspond  to the neck radius equaling to the nuclear radius constant.}
\label{218Th34}
\end{figure}

In $^{240}$Pu, the fission path goes directly from the saddle point to the
asymmetric  fission valley in $^{240}$Pu while it is not the case in $^{218}$Th
where the system from the 3rd minimum at $q_2\approx 1.2$ has a much smaller
barrier towards symmetric fission ($< 0.5$ MeV) than in asymmetric one, where
the barrier is slightly higher ($\approx 1$ MeV) and thicker. It means that
$^{218}$Th nucleus prefers the symmetric fission, which is confirmed by the
experimental yield. In $^{222}$Th the situation is similar, while beyond
$^{226}$Th the path leading to the asymmetric fission begins to be preferred. To
better understand this process, one has to study the PES's in the full 3D 
deformation space. In Fig. \ref{218Th34} the ($q_2,q_3$) cross-sections of the
PES for $^{218}$Th corresponding to different elongations ($q_2=$1.8, 2.0, 2.2,
and 2.3) are shown.
\begin{figure}[t!]
\includegraphics[width=0.95\columnwidth]{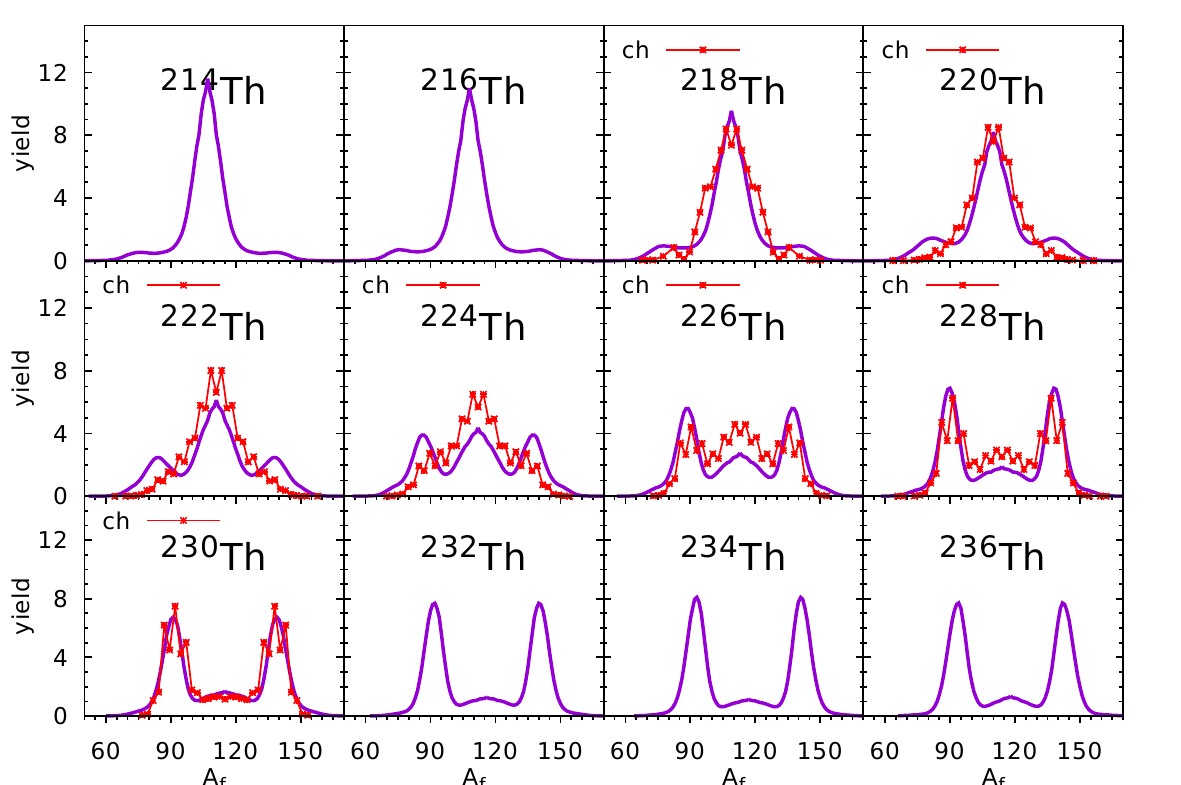}
\caption{Fission fragment mass yields of Th sotopes reproduced by using
constants $E_0$ and $d$ fitted to the experimental data (red stars) taken
from Refs.~\cite{Sch00,Chat19}.}
\label{Th-fit}
\end{figure}

Two minima, one corresponding to the symmetric ($q_3=0$) and the other to the
asymmetric ($q_3\approx 0.12$) configuration, are visible in each cross-section.
At $q_2=1.8$ they are separated by a 4 MeV high barrier which becomes smaller
with growing elongation $q_2$, reaching finally 0.5 MeV height at $q_2=2.1$. At
such elongations, the transition between the symmetric and asymmetric fission is
possible. Both fission valleys are well separated again at the largest
deformations close to the scission line (red line in the figure). So, the Th
nuclei make the ''decision'' {\it where to go} pretty early, i.e., at an early
stage far before the scission configuration. It means that one has to modify the
adjustable parameters $E_0$ and $d$ in order to better describe the transition
between the symmetric and asymmetric fission modes observed in Th nuclei when
the neutron number grows. The new fit performed to the data for Th isotopes only
gives $E_0$=1.5 MeV and $d=$2.5 fm. The resulting mass yields are compared in
Fig.~\ref {Th-fit} with the experimental data. This time the agreement is much
more satisfactory. The new value of the neck parameter $d$ is larger than that
adjusted to all nuclei. It suggests that in Th nuclei, the choice of the
preferable fission mode is made at a thicker neck, i.e., in a pretty early
stage. The smaller value of $E_0$ used for Th isotopes is probably related to
the competition between the symmetric and asymmetric minima.

\section{Summary and conclusions}

In order to briefly summarize our investigations, we can write:
\begin{itemize}
\item The overall accordance of the theoretical FMY estimates with the  
      experimental data indicates that the mac-mic model with the LSD energy 
      for the macroscopic smooth part and the shell and pairing corrections
      evaluated on the basis of the Yukawa-folded single-particle potential
      describes well the potential energy surfaces of actinide nuclei,
\item Three-dimensional set of the Fourier deformation parameters used to
      describe the shape of fissioning nuclei are fully capable to produce a
      wide variety of the shapes of nuclei on their way to fission,
\item The collective 3D model based on the Born-Oppenheimer approximation and
      comprising elongation, mass asymmetry, and neck modes reproduces well 
      the mains features of the fission fragment mass yields data,
\item The Wigner function used to approximate the probability distribution
      related to the neck and mass asymmetry degrees of freedom simmulates in 
      a proper way this distribution for low-energy fission,
\item A neck-breaking probability depending on the size of the neck has to be
      introduced to improve the accordence of our FMY estimates with the
      experimentally measured values.
\end{itemize}

Our mac-mic model and the collective 3D approach, which couples fission mode,
neck, and  mass asymmetry collective vibrations, can describe the main features
of the  fission process in actinide nuclei. The estimated fission barrier
heights deviate not much from their experimental values. The measured fission
fragment mass yields are also reproduced in a satisfactory way. On the other
hand, one has to treat the presented  collective model as a kind of rough tool
which allows to obtain the FMY by a relatively quick calculation. To get more
precise results, one has to use more advanced models in which the whole fission
dynamics and the energy dissipation will be  taken into account. Such
calculations may use the Langevin dynamics (conf. Ref.~\cite{KPo12}) or the
improved quantum molecular dynamics model (ImQMD). The latter method has been
successfully applied to describe the fission process in the heavy ion induced
fission reactions, where the excitation energy increases, leading possibly to a
shorter fission time scale and even to the occurrence of a ternary fission
\cite{QHW19,WGD20}.

The Langevin type calculations, profiting of the PES generated in a mac-mic
approach together with the 3D Fourier shape parametrization as well with the use
of the self-consistent method, are carried out in parallel by our group. 
\vspace{0.5cm}

\noindent
{\bf Acknowledgments}

The authors would like to thank Christelle Schmitt and Karl-Heinz Schmidt for 
supplying us with a part of the experimental data.

\end{document}